\documentclass[11pt]{article}

\usepackage[T1]{fontenc}              
\usepackage[latin9]{inputenc}         

\usepackage[a4paper]{geometry}        

\usepackage{url}

\usepackage{booktabs}

\usepackage{graphicx} 
\usepackage{amssymb,amsfonts,amsmath}

\usepackage{natbib}
\usepackage{mathpazo}
\usepackage{microtype}

\usepackage{color}

\usepackage{todonotes}
 
\newcommand\bX{\boldsymbol X}
\newcommand\bmu{\boldsymbol \mu}
\newcommand\bx{\boldsymbol x}
\newcommand\bw{\boldsymbol w}
\newcommand\bzeta{\boldsymbol \zeta}

\newcommand\expect{\text{E}}
\newcommand\var{\text{Var}}
\newcommand\prob{\text{Pr}}
\newcommand\diag{\text{diag}}

\newcommand{\MALDIquant}{{\tt MALDIquant}}
\newcommand{\binda}{{\tt binda}}
\newcommand{\crossval}{{\tt crossval}}

\newcommand{\figcite}[1]{Fig.~\textbf{\ref{#1}}}
\newcommand{\eqcite}[1]{Eq.~\textbf{\ref{#1}}}

\definecolor{revision}{gray}{0} 

\begin{document}

\date{27 February 2015; revised 30 April 2015}

\title{
Differential Protein Expression and Peak Selection\\ in
Mass Spectrometry Data \\ by Binary Discriminant Analysis
 \\
}

\newcommand\myaddress[2][]{\relax
    {\noindent{\small$^{#1}$#2}}}

\author{Sebastian Gibb$^{1}$ 
and Korbinian Strimmer$^{2}$ 
\thanks{To whom correspondence should be addressed. Email: {\tt k.strimmer@imperial.ac.uk  } 
}
}

\maketitle

\myaddress[1]{Anesthesiology and Intensive Care Medicine, 
  University Hospital Greifswald, Ferdinand-Sauerbruch-Stra\ss{}e,
  D-17475 Greifswald, Germany.\\}
\myaddress[2]{Epidemiology and Biostatistics, School of Public Health,
  Imperial College London, Norfolk Place, London W2 1PG, UK.}
\newpage

\begin{abstract} 

\noindent\textbf{Motivation:} 
Proteomic mass spectrometry analysis is becoming routine in clinical diagnostics, for example to monitor cancer biomarkers using blood samples.  However, differential proteomics and identification of peaks relevant for class separation remains challenging.

\noindent\textbf{Results:} 
Here, we introduce a simple yet effective approach for identifying differentially expressed proteins using binary discriminant analysis.  This approach works by data-adaptive thresholding of protein expression values  and subsequent ranking of the dichotomized features using a relative entropy measure.  Our framework may be viewed as a generalization of the `peak probability contrast' approach of  \citet{TibshiraniHastie+2004} and can be applied both in the two-group and the multi-group setting. 

Our approach is computationally inexpensive and shows in the analysis of a large-scale drug discovery test data set equivalent prediction accuracy as a random forest.  Furthermore, we were able to identify in the analysis of mass spectrometry data from a pancreas cancer study biological relevant and statistically predictive marker peaks unrecognized in the original study.

\noindent\textbf{Availability:}  The methodology for binary discriminant analysis is implemented in the R package  {\tt binda}, which is freely available under the GNU General Public License (version 3 or later) from CRAN at URL \url{http://cran.r-project.org/web/packages/binda/}. 
R~scripts reproducing all described analyzes are available from the web page \url{http://strimmerlab.org/software/binda/}.

\noindent\textbf{Contact:}  \url{k.strimmer@imperial.ac.uk}
\end{abstract}

\newpage

\section{Introduction}

Mass spectrometry, a high-throughput technology commonly used in proteomics, enables the measurement of the abundance of proteins, metabolites, peptides and amino acids in biological samples. The study of changes in protein expression across subgroups of samples and through time provides valuable insights into cellular mechanisms and offers a means to identify relevant biomarkers, e.g, to distinguish among tissue types, or for predicting health status. In practice, however, there still remain many analytic and computational challenges to be addressed, especially in clinical diagnostics \citep{LDM2013}.

A recent overview of statistical issues in the analysis of proteomics mass spectrometry data is \citet{Morr2012} who discusses a wide range of methods ranging from data preprocessing, i.e. removal of systematic bias,  peak identification, peak alignment and quantification and calibration of relative peak intensities, to methods for high-level statistical analysis, such as peak ranking and classification. Of particular importance is the problem of differential protein expression and the identification of peaks informative for group separation and class prediction.   

A special characteristic of mass spectrometry data is their  dual-valued nature, i.e. they contain  both continuous as well as discrete information.   Specifically, a protein may be differentially expressed if its intensity of expression varies among groups and is relatively up- or down-regulated, or if a corresponding peak is either absent or present in a specific group.  Consequently, mass spectrometry intensity matrices typically contain very large amounts of missing values, which renders application of standard statistical methodology from other omics platforms, such as regularized $t$ scores, difficult and potentially suboptimal.  Accordingly, this has initiated the development of new statistical methodology \citep{TibshiraniHastie+2004,WASD2012}.

Two main strategies to address this issue in the high-level analysis of mass-spectrometry data have emerged:
\begin{enumerate}
\item All data are treated as continuous, with missing intensity values set to zero or imputed. Subsequently, standard omics methods are employed, such as $t$-scores for feature selection \citep[e.g.][]{DD2006}.
\item The absence-presence data is used for data analysis in conjunction with the intensity values. \citet{TibshiraniHastie+2004} propose peak probability contrasts (PPC), the absolute difference in frequency of occurrence of a peak, for ranking and feature selection, and also use PPC to improve absence-presence data by dichotomization of intensity values. \citet{WASD2012} propose a test based on the PPC statistic and propose to apply joint FDR control of the union of intensity-based and PPC-based rankings. 
\end{enumerate}

Here, we follow the second route and propose a novel coherent model for differential protein expression and prediction based on binary discriminant analysis. Our approach may be viewed as a generalization of \citet{TibshiraniHastie+2004} and comprises the following:
\begin{itemize}
\item The binary absence-presence data are explicitly modeled by a multivariate Bernoulli distribution.
\item Binary multi-group discriminant analysis (BinDA) is employed for feature ranking, variable selection and prediction. 
\item For ranking of peaks the natural relative entropy variable importance measure coherent with BinDA is used, rather than PPC.
\item Likewise, for dichotomization of the intensity data matrix containing missing values we employ the same entropy-based criterion.
\end{itemize}
As a result, we obtain simple principled framework for analyzing dual-valued mass spectrometry data without the need for imputation,  with a natural measure for variable ranking and for differential protein expression, and with coherent prediction rules.  In contrast to many other methods this approach also allows multiple groups as response variable, and thus extends beyond simple pairwise comparisons. 

The remainder of the paper is structured as follows. Next, we describe in detail the statistical methodology underlying BinDA.  Then, for validation we investigate the performance of the proposed approach in comparison with a random forest on a large-scale chemometric data set. Subsequently, we present a detailed case study analyzing mass spectrometry data from a pancreas cancer study.  For reproducibility, we provide the R package \binda\ implementing our approach and R scripts for all analyzes described.  Finally, we  discuss applicability of the BinDA approach to other molecular data as well as further extensions.

\section{Methods}

\subsection{Setup and notation}

Our analysis starts after the raw mass spectrometry data have been adequately preprocessed, i.e. transformed, smoothed, background-removed, calibrated, aligned, and peak-extracted \citep[e.g.][]{Morr2012,GibbStrimmer2012}.   

We denote the resulting peak intensities by $z_{ij} \geq 0$, with spectrum index $i \in \{1, \ldots , n\}$ and peak index $j \in \{1, \ldots , d\}$.  The data matrix $z_{ij}$ typically contains missing values as not all of the registered $d$ peaks will be present in all of the $n$ spectra.  In a classification setting each spectrum $i$ also carries a class label $y_i  \in \{1, \ldots, K\}$ that assigns it to one of $K$ different groups, for instance health status, tissue type, or treatment outcome.  The label is known for training data and unknown for test data. The sample size in group with label $y$ is $n_y$ with $n=\sum_{y=1}^K n_y$.

From the continuous data $z_{ij}$ we obtain binary peak intensities $x_{ij}$  by thresholding at peak-specific levels  $\bw = (w_1, \ldots, w_d)$.  Specifically, we set $x_{ij}=1$ if the peak is present in sample $i$ and  $z_{ij} \geq w_j$.  Conversely, if the peak is absent or $z_{ij} < w_j$ then $x_{ij} =0$.  The methodology we present here uses the binary matrix $x_{ij}$ for prediction and variable ranking, rather than the original data $z_{ij}$, and it also estimates the thresholds $\bw$.

\subsection{Modeling binary data}

Stochastic models for multivariate binary data are well established \citep[e.g.][]{DaiDingWahba2013,Cox1972}.
A univariate binary random variable $X \sim Be(\mu)$ with two states x=0 and x=1 is completely described by a Bernoulli distribution $Be(\mu)$ with expectation $\expect(X) = \mu$ and variance $\var(X) = \mu (1-\mu)$. 

In the multivariate case this generalizes to  $\bX = (X_1, \ldots, X_d) \sim Be_d(\bmu, \bzeta) $ where $d$ denotes the dimension of the  multivariate Bernoulli (MVB) distribution, $\bmu = (\mu_1, \ldots, \mu_d)$ is the vector of expectations $\expect(\bX) = \bmu$, and $\bzeta$ contains the $2^d -d-1$ interaction parameters.  As in the univariate case the variances $\var(X_j) = \mu_j (1-\mu_j)$ are fully determined by the means. 

In many cases it is useful to ignore the dependencies among the individual variables $X_j$ in order to reduce the number of parameters in the model.  Despite, or perhaps because, of its simplicity the independence ``naive Bayes'' assumption can be very effective, especially for prediction in high dimensions and small sample size, see \citet{HY2001,Park2009}.  For MVB with independent predictor variables the joint probability mass function is given by
\begin{equation*}
 \prob(\bx) = \prod_{j=1}^d \;\begin{cases}
 1-\mu_j & \text{if $x_j = 0$,}
\\
 \mu_j &\text{if $x_j = 1$.}
\end{cases} \\
\end{equation*}
with diagonal covariance matrix  $\var(\bX) = \diag\{ \mu_j (1-\mu_j)\}$.

\subsection{Discriminant analysis with binary predictors}

For prediction of the class associated with an unlabeled spectrum we need to construct a prediction rule.  Here we employ a Bayesian prediction rule similar as in diagonal discriminant analysis (DDA) that is routinely and successfully used, e.g., in transcriptomics \citep{THNC03}.  We call this approach binary discriminant analysis (BinDA).

We first define group-specific models for each group with label $y$,
\begin{equation}
 \prob( \bx | y) = \prod_{j=1}^d \;
\begin{cases}
 1-\mu_{yj} & \text{if $x_j = 0$,} \\
 \mu_{yj} &\text{if $x_j = 1$.}
\end{cases}
\label{eq:condmvb}
\end{equation}
For each group $y$ we also specify a prior probability $\prob(Y=y) = \pi_y$ with $\sum_{y=1}^K \pi_y=1$.  By $\mu_{0j} = \sum_{y=1}^K \pi_y \mu_{yj}$ we denote the pooled mean for each variable $j$, i.e. the mean we would assign if there was only a single category.

The posterior probability of each group is then given by Bayes' theorem
$ \prob(y | \bx) = \prob( \bx | y) \pi_y   /  \prob( \bx )   
$
which after taking the logarithm yields the
discriminant function 
\begin{equation} \color{revision}
d_y(\bx) = \log \prob(y | \bx)  =  \log \pi_y +  \log \prob( \bx | y) + C \,.
\label{eq:mvbpredrule}
\end{equation}
As the purpose of $d_y(\bx)$ is only to compare among different groups we can drop all terms that do not depend on $y$, such as $\prob( \bx )$, represented above by the constant $C$.  Prediction of a label for test data $\bx$   is carried out by choosing the group $y$ that maximizes the discriminant function,
$$
\hat y  = \underset{y}{\arg \max} \; d_y(\bx)
$$
This MVB independence prediction rule has shown to be highly effective \citep[e.g.][]{Park2009}, even if there is correlation among predictors.

Typically, the parameters of discriminant function are unknown themselves and have to be learned themselves from training data, i.e. from spectra with known group labels. The training is done by estimating the means $\mu_{yj}$ and the group probabilities $\pi_y$ in  \eqcite{eq:condmvb} and \eqcite{eq:mvbpredrule}.  We suggest a flexible estimation strategy by employing maximum likelihood estimation for large sample size, and otherwise using regularized estimation. For instance, to estimate the group probabilities we use observed frequencies $\hat\pi_y = n_y/n$ if $n$ is large, and for small $n$ the Stein-type shrinkage estimator of proportions described in \citet{HS09}.

\subsection{Variable ranking and selection}

Closely tied in with prediction is the question which variables are most important for successful assignment of a class label, and, conversely, which variables are irrelevant.  Especially in large-dimensional problems it is very important to remove the null features as the build-up of random noise from these variables can substantially degrade the overall prediction accuracy \citep[cf.][]{AS2010}.  

For ranking features in discriminant analysis with binary variables there have been many, in part contradictory, propositions.  For the case of $K=2$ groups the following criteria, among others, have been used:
\begin{itemize}
\item  The chi-square statistic of independence between response and predictors \citep{AWG2013},
\item  peak probability contrasts $|\mu_{y1} - \mu_{y2}|$  \citep{TibshiraniHastie+2004},
\item Quinlan's information gain measure \citep{BenderMussa+2004}, and
\item ratio of between-group and within-group covariance \citep{WilburGhosh+2002}.
\end{itemize}
See \citet{TKS2004} for many other proposals for measuring associations between categorical outcomes and binary variables.
Only some of the criteria above can also be applied to the multiple group case ($K>2$).

We use a principled approach to variable ranking relying on predictive information, see \citet{GelmanHwangVehtari2014} for an overview.   Conceptually, we use the expected log-predictive density as measure of model fit, and compare the fully specified joint model containing all predictors and the response with a ``no-effects'' model where the response is independent of the predictors.  The difference of expected log-likelihood between full and ``no-effects'' model is given by the relative entropy or Kullback-Leibler divergence $D= KL(F_{\text{full}} || F_{\text{no-eff}})$. The relative contributions of each individual predictor to $D$ then provides a measure of variable importance.  This procedure applied to linear regression with independent predictors results in squared marginal correlations, and applied to diagonal discriminant analysis it yields squared $t$-scores, both of which are optimal measures for variable ranking in their respective settings \citep{ZS2011}.

For independent binary predictors the joint full model is
\begin{equation*}
 \prob( \bx , y)_\text{full} =     \prod_{j=1}^d \;\begin{cases}
 (1-\mu_{yj}) \pi_y & \text{if $x_j = 0$,}
\\
 \mu_{yj} \pi_y &\text{if $x_j = 1$}
\end{cases}
\label{eq:fullmodel}
\end{equation*}
whereas the no-effects model is 
\begin{equation*}
 \prob( \bx , y)_\text{no-eff} = \prod_{j=1}^d \;\begin{cases}
 (1-\mu_{0j}) \pi_y & \text{if $x_j = 0$,}
\\
 \mu_{0j} \pi_y &\text{if $x_j = 1$.}
\end{cases}
\label{eq:noeffmodel}
\end{equation*}
This results in 
\begin{equation}
\begin{split}
D & =\sum_{j=1}^d \sum_{y=1}^K \; \left(  \mu_{yj} \pi_y  \log\left(\frac{\mu_{yj}}{\mu_{0j}}\right)
 + (1-\mu_{yj})\pi_y  \log\left(\frac{1-\mu_{yj}}{1-\mu_{0j}}\right)      \right) \\
 & = \sum_{j=1}^d \sum_{y=1}^K \pi_y \; KL\left( Be(\mu_{yj}) ||  Be(\mu_{0j}) \right) \\
& \approx  \sum_{j=1}^d \frac{1}{2}  \sum_{y=1}^K \pi_y \;  
  \left( {  \mu_{yj}-  \mu_{0j}   \over \sigma_j } \right)^2  = \sum_{j=1}^d S_j
\end{split}
\label{eq:entropyscore}
\end{equation}
where $\sigma_j^2 =\mu_{0j} (1-\mu_{0j}) $ is the variance of $Be(\mu_{0j})$.
For the special case of $K=2$ groups $S_j$ simplifies to
$$
S_{j (K=2)} =  \frac{ \pi_1 \pi_2}{2} \;   \left( {  \mu_{1j}-  \mu_{2j}   \over \sigma_j } \right)^2.
$$
By construction,  the score $S_j$ is a measure of variable importance of feature $j$ where $S_j$ is a weighted sum of the squared $z$-scores that compare each group mean with the overall pooled mean. This is precisely analogous to the pooled-mean formulation of discriminant analysis described in \citet{AS2010}.  If the variances $\sigma^2_j$ are similar across features, then $S_{j (K=2)}$ is apart from a scale factor the squared peak probability contrast.

As above for learning the discriminant function, we use for estimation of the entropic ranking scores $S_j$  either maximum likelihood or shrinkage estimates of proportions, depending on sample size.  

Noting that $n  \hat\pi_y / (1-\hat\pi_y) = (1/n_y - 1/n)^{-1}$ we may also introduce squared $t$-scores 
$$t^2_{y0} = n  \frac{\hat\pi_y}{1-\hat\pi_y}  
\left( {  \hat\mu_{yj}-  \hat\mu_{0j}   \over \hat\sigma_j } \right)^2 
$$
that are properly scaled to allow to contrast the individual contributions of each class relative to each other, similar as in the methods described in \citet{AS2010} and \citet{THNC03}.  The estimated ranking score may also be expressed in terms of a weighted  sum of the $t$-scores via 
$$
\hat{S}_j = \sum_{y=1}^K (1-\hat\pi_y) t^2_{y0} / (2n) \,.
$$
After ranking variables  according to the estimated scores $\hat{S}_j$, with highest scores indicating the most relevant predictors, we use cross-validation to evaluate prediction accuracy for different numbers of included predictors to determine a suitable cutoff. 

\subsection{Dichotomization}

With the above setup for BinDA it is straightforward to perform dichotomization. Specifically, we choose thresholds $\bw = (w_1, \ldots, w_d)$ to discretize the continuous data $z_{ij}$  to maximize the entropy score $D$ (\eqcite{eq:entropyscore}).  As in our model assume the predictors are assumed to be independent we can optimize each threshold $w_j$ independently by maximizing the individual~$S_j$.  {\color{revision} Note that the same entropy measure is used both for determining the thresholds and for ranking the predictors, thus ranking and discretization is done in an integrative fashion.}

\section{Results}

\subsection{Implementation and reproducible research}

We have implemented our approach for  multi-class discriminant analysis using binary predictors including functions for  variable ranking and dichotomization in the R package \binda{} that is freely available under the GNU General Public License (version 3 or later) from URL \url{http://cran.r-project.org/web/packages/binda/}.  For reproducibility of the analyzes presented in this paper we provide corresponding R scripts at \url{http://strimmerlab.org/software/binda/}.

\subsection{Validation of \binda}

Binary discriminant analysis (BinDA) has been studied extensively and is well established in the literature \citep[e.g.][]{Cox1972}. More recently, it was demonstrated that BinDA with naive Bayes assumption can yield high rates of predictive accuracy even if the underlying assumption of independence of predictors is not met  \citep{AWG2013,Park2009, BenderMussa+2004,WilburGhosh+2002}.

For validation of our implementation of BinDA in the R package \binda{} we analyzed a large-scale chemometric test data set.  Specifically, we investigated the  ``Dorothea'' drug discovery data set from the NIPS 2003 feature selection challenge \citep{Guyonetal2005}. The data set contains $d=100,000$ binary features describing the three-dimensional properties of chemical compounds that either bind  (response label +1) or not (label -1) to thrombin, an enzyme involved in blood clotting.

We  used the ``Dorothea'' training data with $n_\text{train}=800$ samples and corresponding class labels to learn a classifier with \binda.  Subsequently,  we applied the resulting classification rule  to the validation data set with $n_\text{val}=350$ samples and predicted the sample labels of the validation data.  As for the validation data the true labels are known we were then able to compute the actual prediction accuracy, i.e. the proportion of correctly identified labels. 

{\color{revision} Due to its algorithmic simplicity  training the classifier and ranking variables with \binda{} was computationally inexpensive. Applying class-balanced 5-fold cross-validation with 20 repetitions using the R package \crossval{}  on the training data alone we determined that the response can be predicted well with only very few top ranking predictors included in the classification rule. For instance, a \binda{} classifier with 3 predictors yielded prediction accuracy on the validation set of 0.9371 and of 0.9429 if 10 predictors were used.}  The predictive accuracy without any variable selection including all 100,000 predictors was 0.9057.

For comparison we also trained a random forest \citep{Breiman2001}, a tree-based machine learning approach  that emerged as one of the overall best performing methods for classification in a recent systematic study \citep{FCBA2014}. Due to the high-dimensionality the running time for learning the random forest from the training data was two magnitudes slower than \binda, {\color{revision} taking 652 seconds on our workstation in comparison to 5 seconds for \binda.}  The random forest yielded an accuracy of 0.94 for prediction of the labels of the independent validation data set.  This analysis confirms that BinDA, though very simple, is able, at least for this data,  to perform prediction as accurately as random forest.
{\color{revision} Correspondingly, if variables in the random forest were ranked according to the Gini variable importance measure the top-ranking features were mostly identical to those ranked best by BinDA, which indicates that BinDA is indeed able to select the most relevant variables.}

\subsection{Analysis of pancreas cancer proteomics data}

\subsubsection{Pancreas cancer study}

For illustration of our proposed approach to classification and peak ranking in mass spectrometry data we also reanalyzed experimental proteomics data from a pancreas cancer study conducted in Leipzig and Heidelberg \citep{FiedlerLeichtle+2009}. For the training data set of this study 40 patients with diagnosed pancreas cancer as well as 40 healthy controls  were recruited. Each participant of the study donated serum samples which provided the basis for MALDI/TOF measurements.   For each sample 4 technical replicates were obtained. Due to the presence of strong batch effects in our analysis below we restrict ourselves to patients and controls from Heidelberg, leading to a raw data set containing 160 spectra for 40 probands, 

The aim of the study was to determine biomarkers to discriminate patients with pancreas cancer from healthy persons. \citet{FiedlerLeichtle+2009} found marker peaks at m/z 3884 (double charged) and 7767 (single charged) and correspondingly supposed platelet factor 4 (PF4) as potential marker, arguing that PF4 is down-regulated in blood serum of patients with pancreatic cancer. 

\subsubsection{Preprocessing and dichotomization}

For preprocessing of the raw mass spectrometry data we employed the standard analysis pipeline implemented in the  R package \MALDIquant{} \citep{GibbStrimmer2012}.  Specifically, the raw data were variance-stabilized, smoothed, baseline-corrected, TIC-standardized, and aligned. Technical replicates were then averaged, peaks were identified and corresponding intensities extracted from each averaged spectrum.  Precise details on the preprocessing can be found in the  R script. As a result a protein expression matrix of size  40 patients times 166 peaks was obtained. In total 26\% of intensities in the matrix were missing, corresponding to about 44 missing peaks per spectrum.

\begin{figure}
\begin{center}
\includegraphics[width=\linewidth]{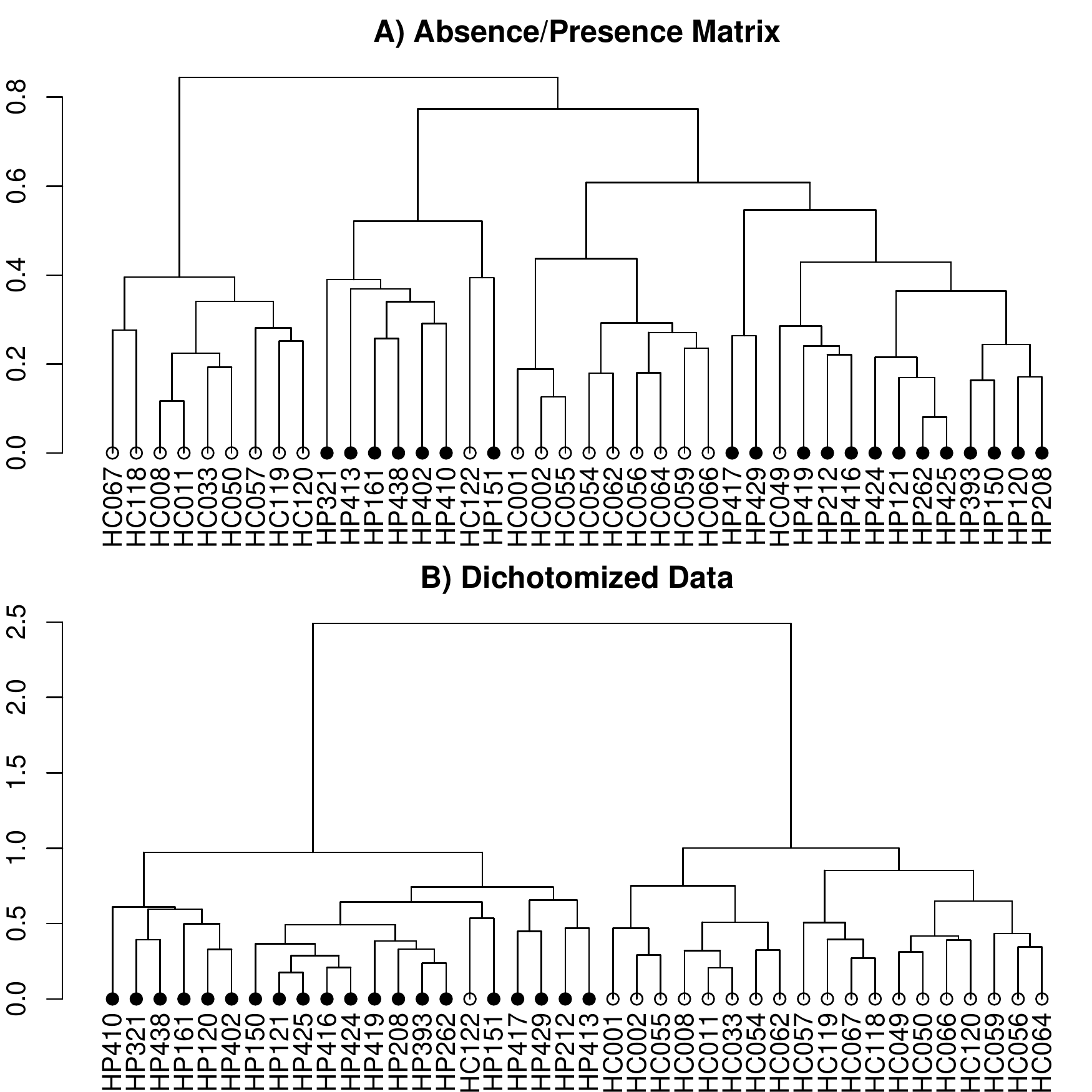}
\end{center}
\caption{Clustering of  samples from the pancreas cancer study of \citet{FiedlerLeichtle+2009} using (A) the original absence-presence data and (B)  the optimized binary matrix. Filled circles indicate pancreas cancer samples, empty circles healthy controls. { \color{revision} For clustering we employed Ward's agglomerative hierarchical clustering based on a Jaccard distance matrix computed
using R standard functions {\tt hclust()} with {\tt method="ward.D2"} and {\tt  dist()} with {\tt method="binary"}. }
\label{fig:clustering}} 
\end{figure}

Subsequently, we performed dichotomization of the intensity matrix using the relative entropy criterion of \eqcite{eq:entropyscore}. To illustrate the improvement of the resulting binary data matrix over the original absence-presence matrix we conducted hierarchical clustering on the samples.  As can be seen in \figcite{fig:clustering} the clustering based on the optimized binary matrix almost perfectly separates pancreas cancer samples from healthy samples, indicating that there is a strong signal in the data.

\subsubsection{Peak ranking and differential expression thresholds}

\begin{figure}
\begin{center}
\includegraphics[width=\linewidth]{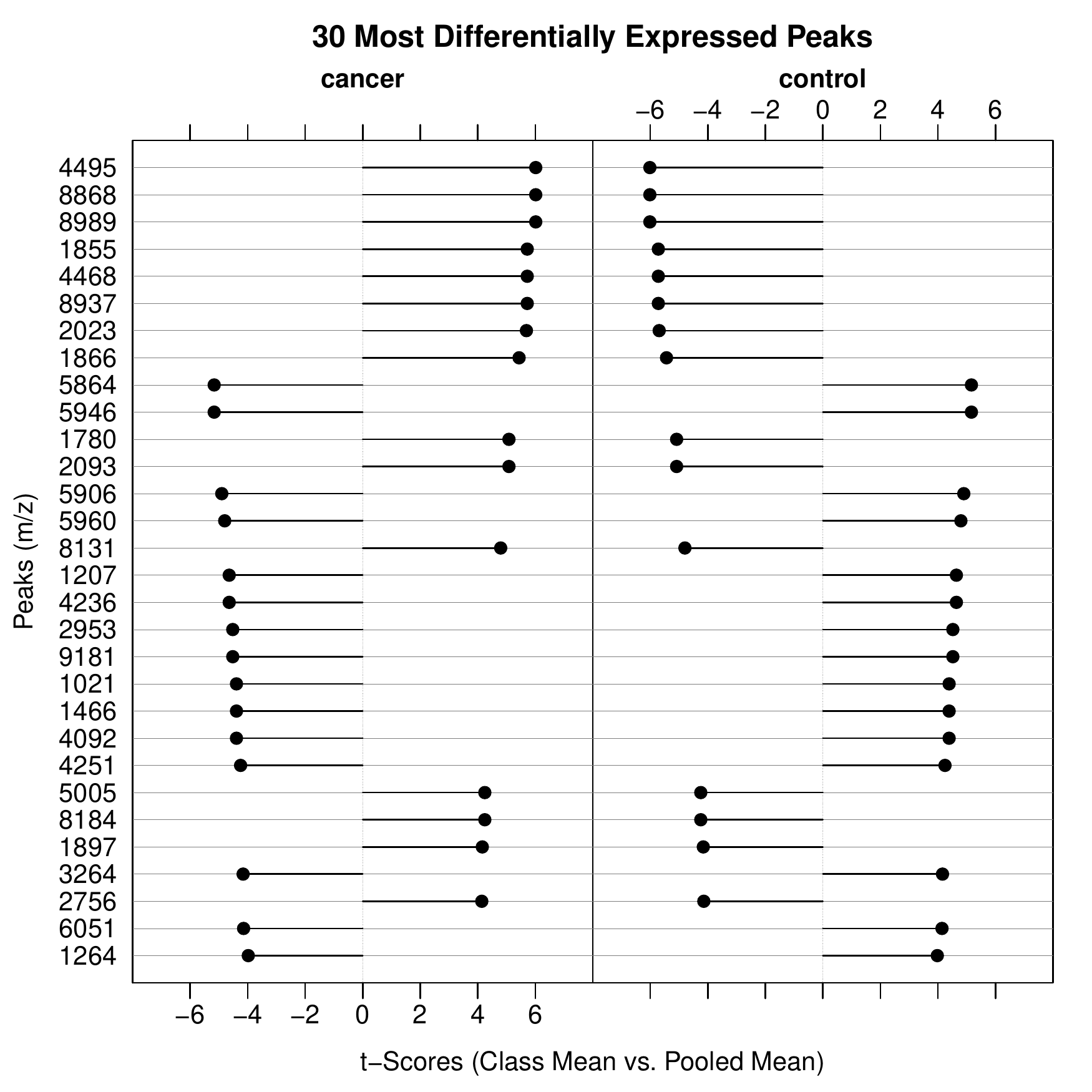}
\end{center}
\caption{Ranking of 166 peaks in the preprocessed spectra from the pancreas cancer study according to BinDA framework. \label{fig:ranking}} 
\end{figure}

In order to identify features responsible for the separation of cancer versus healthy samples in \figcite{fig:clustering}~B we  applied peak ranking for binary data according to BinDA.  The resulting ranking of the 30 best discriminating peaks is shown in \figcite{fig:ranking}. As a consequence of the discrete data, the first three top-ranking peaks with m/z values 4495, 8868, and 8989) achieved the same maximum score, followed by the next three peaks 1855, 4468, and 8937, that also achieved an identical score.

We note that none of these peaks were identified in the original study.   The two PF4 peaks with m/z 3884 and 7768 (the slight difference is due to the MALDIquant alignment procedure) rank on places 148-151 and  157-163, respectively.

Using cross-validation we estimated prediction errors for group separation from the binary data matrix.
{\color{revision} As for the ``Dorothea'' data set we employed class-balanced 5-fold cross-validation with 20 repetitions.}
Interestingly, using only 5 predictors was sufficient to achieve an accuracy of 0.96, sensitivity of 0.96, specificity of 0.97, positive predictive value of 0.97 and negative predictive value of 0.95. This indicates that the observed clear separation between cancer and control samples in \figcite{fig:clustering}~B is attributable to only very few features of the data.

\begin{figure}
\begin{center}
\includegraphics[width=\linewidth]{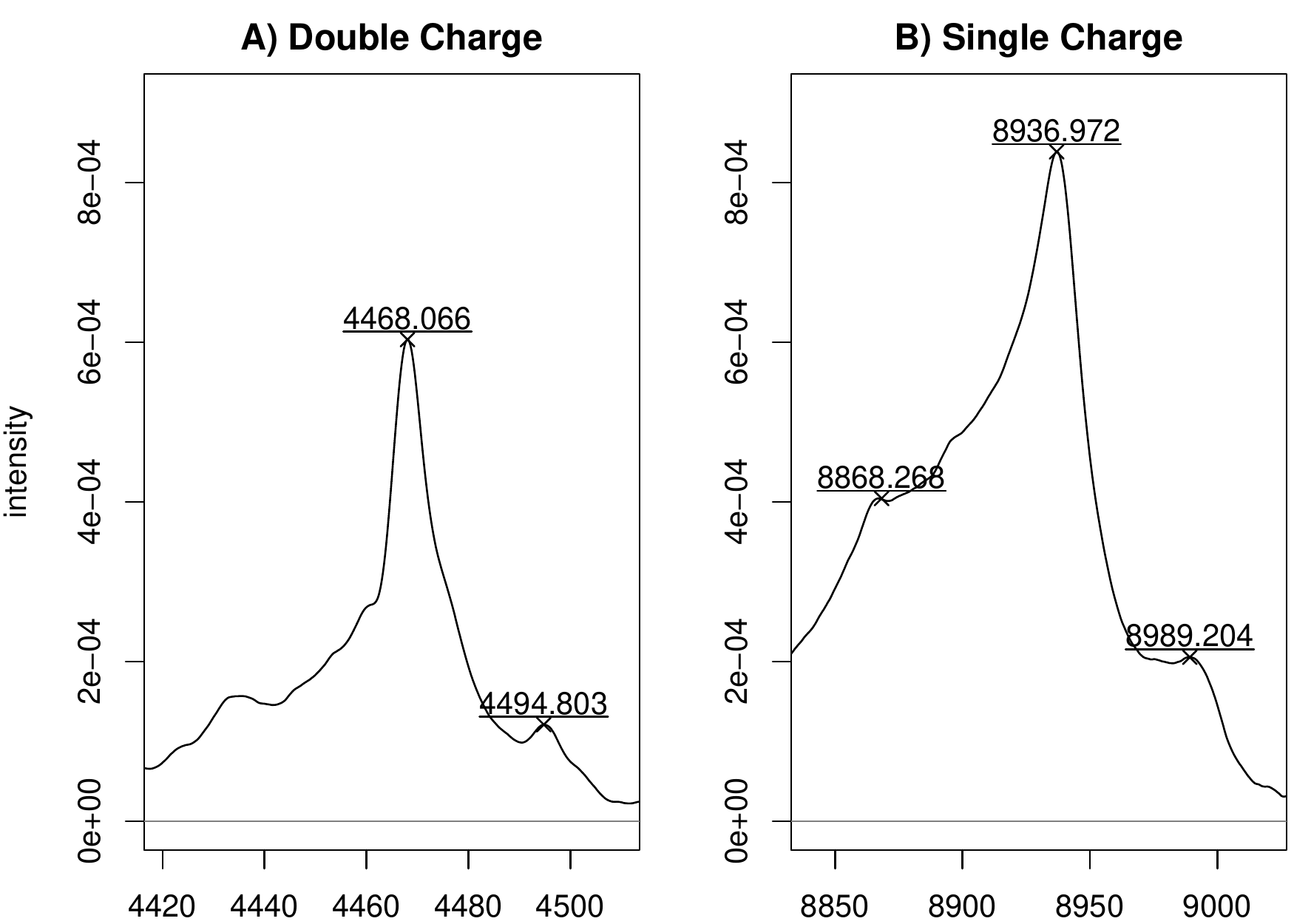}
\end{center}
\caption{Top ranking peak group containing 5 differentially expressed peaks: (A) double- and (B) single-charged peaks. \label{fig:peaks}} 
\end{figure}

Visual inspection of the group of top-ranking differentially expressed peaks revealed a further pattern (\figcite{fig:peaks}). First, five of the peaks are all part of the same peak group.  Second, the peak group appears both in a single charged (m/z 8868, 8937, 8989) version as well as in a mirrored double charge version (m/z 4468 and 4495).  This  affirms that there must a underlying biological marker driving the observed changes between cancer and control samples.

\begin{figure*}
\centering
\includegraphics[width=\linewidth]{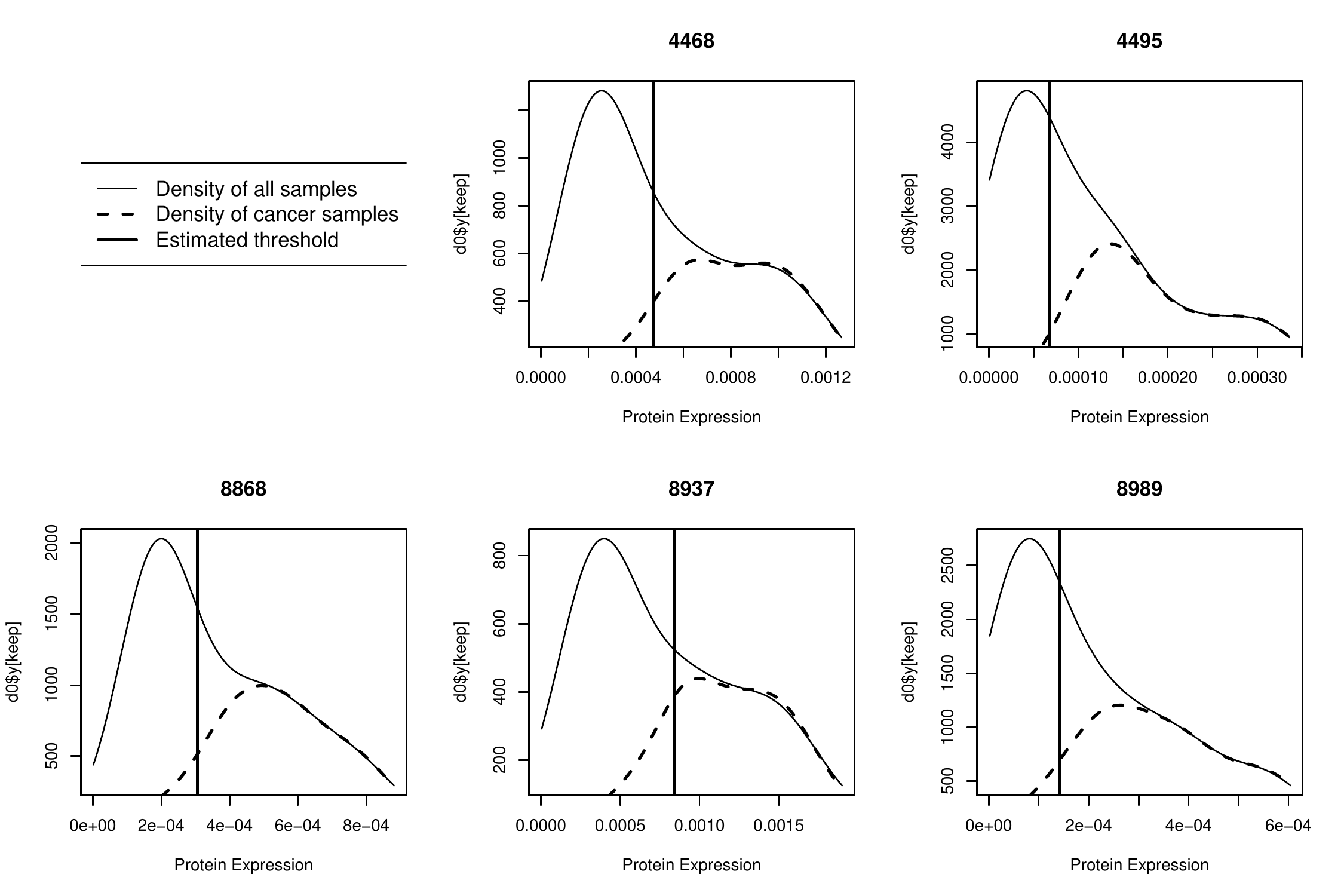}
\caption{Densities of expression values and estimated thresholds for the top ranking peaks. Each column shows the single charged (bottom row) and the corresponding double charged variant (top row). \label{fig:thresholds}} 
\end{figure*}

To study this further, we inspected the intensities for the peaks belonging to the differentially expressed peak group. \figcite{fig:thresholds} shows the overall density, as well as the sub-density for the cancer samples, along with the dichotomization threshold estimated by \binda. For all five peaks the expression respectively the underlying protein abundance is up-regulated in the cancer samples compared with the controls.  In addition, the estimated thresholds provide an effective means to separate the two groups.

\subsubsection{Biological relevance}

Finally, we also tried to identify the biological molecules behind the differentially expressed peak group shown in \figcite{fig:peaks}. Specifically, we used the TagIdent tool \citep{GH+2005} with settings Mw 8936.97, Mw range 0.05\% and organism {\it homo sapiens} to query the UniProtKB/Swiss-Prot data base \citep{uniprot2015}.  This indicated a potential link of the central peak m/z 8937 to PDPFL\_HUMAN, the pancreatic progenitor cell differentiation and proliferation factor-like protein, as well as to a fragment of C3adesArg, an acylation stimulating protein.  The increased abundance of PDPFL\_HUMAN in pancreas cancer tissue appears highly plausible, and the increased concentration of C3adesArg in serum of cancer patients has also been reported previously \citep[e.g.][]{OK+2011}. 

Another biologically relevant result of our analysis based on BinDA is that the originally proposed PF4 marker is not differentially expressed and hence cannot be used to distinguish between cancer and healthy samples.

\section{Discussion}

We have presented a simple yet effective approach to differential expression and classification for mass spectrometry data using binary discriminant analysis.  Our approach may be viewed as generalization of \citet{TibshiraniHastie+2004} and can be applied also for multi-group discriminant analysis. A particular feature is the use of the same relative entropy criterion for peak ranking and selection and for dichotomization of the continuous protein intensity data.  In addition, we obtain decision thresholds from the protein intensities that are biologically and diagnostically easy to interpret.

In illustrative analysis of high-dimensional drug discovery data we showed that our approach implemented in the R package \binda{} is computationally effective and yet competitive with a random forest.  Furthermore, in reanalysis of proteomics data from a pancreas cancer study we found statistically predictive marker peaks to tumor cell growth unrecognized in the original analysis.   This confirms the importance of reproducible research in proteomics, where it is unfortunately still not common to provide analysis scripts and software openly.

In addition to mass spectrometry analysis, there are many bioinformatics applications in which binary data are collected, and hence in which the present methodology and software will potentially be useful.  Examples include meta-genomics, where the absence and presence of proteins and genes is compared to a pan-genome \citep{MS+2008}, community analysis by DNA fingerprinting \citep{WilburGhosh+2002}, and chemometrics \citep{BenderMussa+2004}. 

Exploring additional applications may also lead to further  methodological extensions of the procedures currently implemented in \binda, such as modeling overdispersion, e.g., by employing the Beta-Bernoulli rather than Bernoulli distribution, and to take account of interactions among predictors, e.g., by modeling pair-wise correlation.

\section*{Acknowledgements}
We thank \citet{FiedlerLeichtle+2009} for kindly providing us with their data. In addition we thank PD Dr. med. Alexander B. Leichtle and Prof. Dr. med.  Martin Fiedler for very helpful discussions.

\newpage

\bibliographystyle{apalike}
\bibliography{preamble,proteomics,genome,stats,array,entropy}

\end{document}